\documentclass{aa}
\usepackage{txfonts}

\usepackage{graphicx}
\begin{document}
\title{WR\,20a: a massive cornerstone binary system comprising two extreme early-type stars\thanks{Based on observations collected at the European Southern Observatory (La Silla, Chile)}}
\author{G.\ Rauw\inst{1}\fnmsep\thanks{Research Associate FNRS (Belgium)} \and M.\ De Becker\inst{1} \and Y.\ Naz\'e\inst{1}\fnmsep\thanks{Research Fellow FNRS (Belgium)} \and P.A.\ Crowther\inst{2} \and E.\ Gosset\inst{1}\fnmsep$^{\star\star}$ \and H.\ Sana\inst{1}\fnmsep$^{\star\star\star}$ \and \newline K.A.\ van der Hucht\inst{3}\fnmsep\inst{4} \and J.-M.\ Vreux\inst{1} \and P.M.\ Williams\inst{5}}
\offprints{G.\ Rauw}
\mail{rauw@astro.ulg.ac.be}
\institute{Institut d'Astrophysique et de G\'eophysique, Universit\'e de Li\`ege, All\'ee du 6 Ao\^ut, B\^at B5c, 4000 Li\`ege, Belgium 
\and Department of Physics \& Astronomy, University of Sheffield, Hicks Building, Hounsfield Rd., Sheffield, S3 7RH, UK \and SRON National Institute for Space Research, Sorbonnelaan 2, 3584 CA Utrecht, The Netherlands \and Astronomical Institute Anton Pannekoek, University of Amsterdam, Kruislaan 403, 1098 SJ Amsterdam, The Netherlands \and Institute for Astronomy, University of Edinburgh, Royal Observatory, Blackford Hill, Edinburgh, EH9 3HJ, UK}
\date{Received date / Accepted date}
\abstract{We analyse spectroscopic observations of WR\,20a revealing that this star is a massive early-type binary system with a most probable orbital period of $\sim 3.675$\,days. Our spectra indicate that both components are most likely of WN6ha or O3If$^*$/WN6ha spectral type. The orbital solution for a period of $3.675$\,days yields extremely large minimum masses of $70.7 \pm 4.0$ and $68.8 \pm 3.8$\,M$_{\odot}$ for the two stars. These properties make WR\,20a a cornerstone system for the study of massive star evolution. 
}
\maketitle
\section{Introduction}
Although it is generally accepted that Wolf-Rayet (WR) stars are evolved descendants of massive O-stars, many open questions remain about the evolution of these objects. Among the most fundamental issues is the importance of mass transfer in the evolution of massive O + O binaries (e.g.\ Vanbeveren et 
al.\ \cite{Vb}). It is therefore of prime interest to achieve as complete a census of WR binaries as possible and to investigate the fundamental properties of individual massive binary systems.  

Among the 227 objects listed in the VIIth Catalogue of Galactic Wolf-Rayet Stars (van der Hucht \cite{vdH}), there are only 52 known spectroscopic binaries and only 32 of them have reliable orbital solutions. In addition to these confirmed binaries, van der Hucht (\cite{vdH}) provides a list of 30 rather faint and poorly studied objects that are potential WR + OB binaries. Most of these binary candidates were classified as such because of their comparatively weak emission lines that could be the result of the dilution by the continuum light of an OB-type companion. In order to search for a composite spectrum - and hence confirm multiplicity - we performed a snapshot spectroscopic survey of a sample of WR binary candidates. One of the most interesting objects that emerged from our survey was WR\,20a.

\object{WR\,20a} (= THA\,35-{\sc ii}-036 = SMSP2) was catalogued by Th\'e (\cite{The}) in an H$\alpha$ emission line survey, and first discovered to be a WR star by Shara et al.\ (\cite{Shara}) in the course of a deep narrow and broadband photometric survey for galactic Wolf-Rayet stars in the southern Milky Way. The latter authors also obtained a low-resolution (2 pixel resolution of 7\,\AA) spectrum of the star and proposed a WN7 spectral type, although they noted the lack of a He\,{\sc ii} $\lambda$\,5411 emission line. Shara et al.\ (\cite{Shara}) suggested that WR\,20a had a rather high H/He abundance ratio. Later, the object was re-classified as WN7:h/WC by Shara et al.\ (\cite{Shara2}). The star is believed to be a member of the young open cluster Westerlund\,2 in the core of the H\,{\sc ii} region RCW\,49 (Moffat et al.\ \cite{Moffat}). 

\begin{figure*}
\begin{center}
\resizebox{15cm}{9.7cm}{\includegraphics{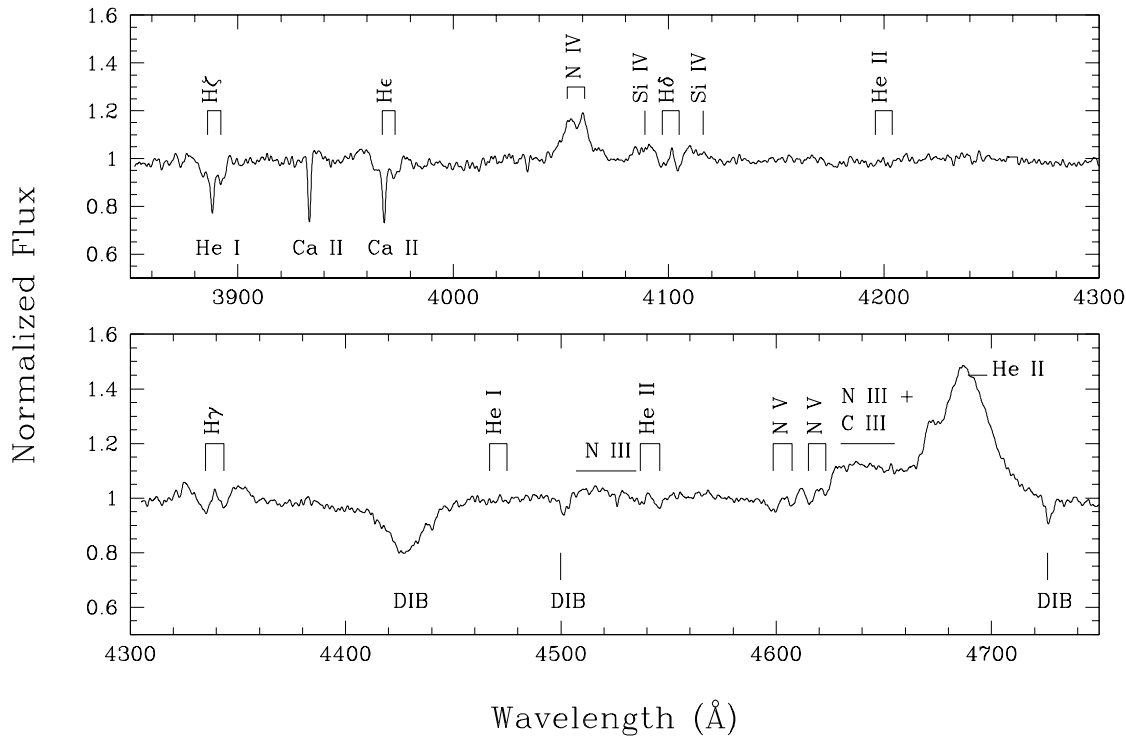}}
\end{center}
\caption{Spectrum of WR\,20a as observed on HJD\,2\,453\,034.715 (violet setting, top panel) and 2\,453\,034.741 (blue setting, lower panel). Note the double N\,{\sc iv} $\lambda$\,4058 emission line as well as the double N\,{\sc v} $\lambda$\,4604 and N\,{\sc v} $\lambda$\,4620 absorptions. The stellar lines are labelled above the spectrum, whereas the interstellar lines and diffuse interstellar bands (DIBs) are indicated below.\label{spec}}
\end{figure*}
In this letter, we first present our blue/violet observations of WR\,20a (Sect.\,\ref{obs}). The spectral features as well as the nature of WR\,20a are revisited in Sect.\,\ref{spect}, whilst Sect.\,\ref{solorb} deals with an analysis of the radial velocities. Finally, we briefly discuss the implications of our results in Sect.\,\ref{discuss}. 
\section{Observations \label{obs}}
Two snapshot spectra of WR\,20a were obtained in March 2002 with the EMMI instrument mounted on ESO's 3.5\,m New Technology Telescope (NTT) at La Silla. The EMMI instrument was used in the BLMD medium dispersion spectroscopic mode with grating \# 3 (1200 lines\,mm$^{-1}$) providing a wavelength coverage of 460\,\AA\ per set-up. The central wavelength was set to 4080\,\AA\ (violet setting) or 4530\,\AA\ (blue setting). The exposure times were 30\,min for each setting and the corresponding S/N was about 100. The spectral resolution as determined from the {\tt FWHM} of the lines of the Th-Ar comparison spectra was about 1.25\,\AA. The data were reduced in the standard way using the {\tt long} context of the {\sc midas} package. The violet spectrum taken on HJD\,2\,452\,355.60 revealed double absorption lines of hydrogen and helium  as well as a moderately strong double-peaked N{\sc iv} $\lambda$\,4058 emission. On the other hand, the blue spectrum obtained on HJD\,2\,452\,354.56 showed single absorption lines of hydrogen, He\,{\sc ii} and N\,{\sc v}. These properties suggested that WR\,20a could be a short period binary consisting of two very early extreme Of-type stars (Driesens \cite{Driesens}). We thus organized a follow-up observing campaign to confirm the binarity of WR\,20a. The star was monitored over 12 consecutive nights in January-February 2004 with the same instrumentation as in March 2002. The weather conditions were good except for the last two nights where the seeing and the transparency were poor. The data were reduced in the same way as in 2002 and the complete journal of the useable observations is given in Table\,\ref{journal}.
\section{Results}
\subsection{The spectrum of WR\,20a \label{spect}}
Our data reveal that WR\,20a is indeed a short-period binary system (see below). All prominent spectral features appear double with nearly identical intensities on most of our spectra (see e.g.\ Fig.\,\ref{spec}) indicating that the components of WR\,20a must have identical spectral types and a visible brightness ratio of $\sim 1$. The spectrum of WR\,20a displays many absorption lines of H\,{\sc i}, He\,{\sc ii} and N\,{\sc v} unlike what we would expect if WR\,20a were indeed a `classical' WN7 star (see however below). Only a few (rather narrow) emission lines are seen in the blue/violet spectrum: N\,{\sc iv} $\lambda$\,4058, He\,{\sc ii} $\lambda$\,4686 as well as a broad complex probably due to N\,{\sc iii} $\lambda\lambda$ 4634-41, C\,{\sc iii} $\lambda\lambda$ 4647-50 and possibly C\,{\sc iv} $\lambda$\,4658. The Balmer absorption lines also display some broad underlying emission wings on several spectra. The emissions seen around the H$\delta$ absorption in Fig.\,\ref{spec} could actually be due to Si\,{\sc iv} $\lambda\lambda$ 4089, 4116, but the spectra obtained on other nights do not allow us to unambiguously confirm this identification.  In addition to the strong diffuse interstellar bands, we note the presence of the interstellar He\,{\sc i} $\lambda$\,3889 absorption line. This latter feature was also observed in the spectra of several early-type stars of the Carina Nebula (Walborn \& Hesser \cite{WH}). This line is not usually seen in interstellar sightlines (cf.\ $\zeta$\,Oph, Shulman et al.\ \cite{Shul}) and may be formed locally as in the case of the Carina stars. 

No stellar He\,{\sc i} lines are clearly seen in the spectrum. Together with the presence of rather strong N\,{\sc v} absorption lines, this would suggest a very early O-type classification for the components of WR\,20a. The presence of He\,{\sc ii} $\lambda$\,4686 in strong emission further suggests a supergiant luminosity class. Comparing our spectra to the classification scheme of Walborn et al.\ (\cite{WHL02}, see also Walborn \cite{Wal}), we would propose an O3\,If$^*$ spectral type for both stars. However, distinguishing an extreme Of star from a weak-lined WNL star is a very tricky issue. In this context, Crowther \& Dessart (\cite{CD}) proposed EW(He\,{\sc{ii}} $\lambda\,4686) = 12$\,\AA\ as a boundary between O3\,If$^*$/WN and WN5-6 stars. For WR\,20a, we determine EW(He\,{\sc{ii}} $\lambda\,4686) = 13.1$\,\AA\ with a 1-$\sigma$ dispersion of $1.3$\,\AA, right at the borderline proposed by Crowther \& Dessart. Since the He\,{\sc ii} $\lambda$\,4686 emission could contain a contribution from a wind interaction process, we consider this an upper limit on the actual strength of the line. Comparing the spectrum of WR\,20a obtained near conjunction with the spectra of several extreme Of and weak-lined WN stars, we find that the components of WR\,20a are indeed intermediate between Mk\,42 (O3If$^*$/WN) and HD\,97950C (= WR\,47c, WN6ha). Finally, the spectrum of WR\,20a presented by Shara et al.\ (\cite{Shara}) indicates that the H$\beta$ line is in weak emission. The latter feature favours a WN6ha spectral type. We shall thus adopt a WN6ha classification for both components of WR\,20a, though we cannot fully exclude an O3\,If$^*$/WN6ha type. The Shara et al.\ (\cite{Shara}) spectrum of WR\,20a suggested the presence of a C\,{\sc iv} $\lambda$\,5808 emission with a normalized intensity stronger relative to He\,{\sc i} $\lambda$\,5876 than in normal WN stars, thus leading to the WN/WC classification proposed by Shara et al.\ (\cite{Shara2}). Note that our blue/violet spectra do not provide support for such a hybrid type.  
\subsection{Orbital solution \label{solorb}}
We have measured the radial velocities (RVs) of the components of WR\,20a by fitting Gaussians. The typical uncertainties on individual RV measurements are of order 10 -- 15\,km\,s$^{-1}$ for those phases when the components are resolved\footnote{Note that we also determined the heliocentric radial velocities of the sharp interstellar features: we obtain RVs of $-16.6 \pm 15.0$, $-7.8 \pm 10.5$ and $-10.2 \pm 14.0$\,km\,s$^{-1}$ for the He\,{\sc i} $\lambda$\,3889, Ca\,{\sc ii} $\lambda$\,3934 and Ca\,{\sc ii} $\lambda$\,3968 lines. The $1\sigma$ dispersions listed here provide some rough indication of the accuracy of our wavelength calibration.}. 

\begin{figure}[htb]
\begin{center}
\resizebox{8cm}{8.2cm}{\includegraphics{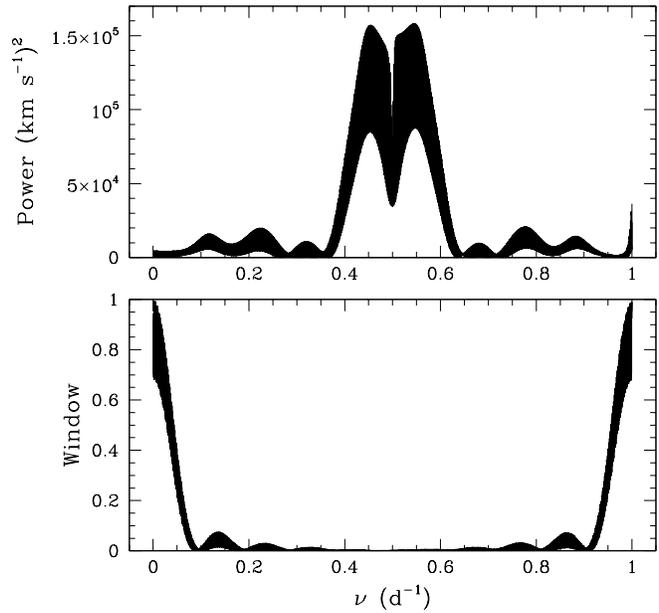}}
\end{center}
\caption{Top panel: power spectrum of our $|RV_1 - RV_2|$ time series of WR\,20a. The bottom panel provides the corresponding power spectral window.\label{fig-period}}
\end{figure}

A major issue in determining the orbital period arises from the fact that the spectral types of both stars are identical. In fact, the spectra do not allow us to distinguish the components {\it a priori}. Therefore, we performed a period search on the absolute values of the radial velocity differences $|RV_1 - RV_2|$. In this way, we expect to find the highest peak in the periodogram at $\nu_1 = 2/P_{\rm orb}$. Using the Fourier technique of Heck et al.\ (\cite{HMM}, see also Gosset et al.\ \cite{Gosset}), we find that the most likely value of $\nu_1$ is $0.544 \pm 0.009$\,day$^{-1}$. However, there is a strong ambiguity with the $1 - \nu_1$ alias at $0.453 \pm 0.009$\,day$^{-1}$ which produces only a slightly smaller (formal) peak in the periodogram (see Fig.\,\ref{fig-period}). Our preferred value for the orbital period is therefore $3.675 \pm 0.030$\,days, but we cannot discard the alternative value of $4.419 \pm 0.044$\,days. Alternatively, if the (slightly smaller) peak at the $1 + \nu_1$ alias ($1.547$\,day$^{-1}$) were to be selected, the orbital period would be $\sim 1.29$\,days, i.e.\ even shorter than the 1.64\,day period of CQ\,Cep, the shortest period WN binary system known to date (see e.g.\ van der Hucht \cite{vdH}). 

As a next step, we determined the apparent systemic velocity of each spectral line from an orbital solution obtained for the RVs of the specific line. In this way, we found apparent systemic velocities of $-75$, $-18$, $-77$, $-42$, $-51$ and $-26$\,km\,s$^{-1}$ for the H$\gamma$, He\,{\sc ii} $\lambda$\,4542, N\,{\sc v} $\lambda$\,4604, N\,{\sc v} $\lambda$\,4620, N\,{\sc iv} $\lambda$\,4058 and H$\delta$ lines respectively. Finally, we averaged the RVs of the different lines listed above after subtracting their systemic velocities. The results are listed in Table\,\ref{journal}. 
\begin{table}
\caption{\label{journal} Journal of our spectroscopic observations of WR\,20a. The first and second columns provide the heliocentric Julian date (in the format HJD$ - 2\,450\,000$) and the orbital phase assuming a period of $3.675$\,days and adopting phase $\phi = 0.0$ as given in Table\,\ref{tab-orbit}. The third column indicates whether the blue (B) or violet (V) spectral range was observed. The recentred and averaged primary and secondary radial velocities are given in columns 4 and 5 respectively (see text). Finally, the last column yields the weight that we assigned to the various RV data in the orbital solution.}
\begin{tabular}{c c c r r c}
\hline\hline
Date & $\phi$ & Set-up & RV$_1$ & RV$_2$ & weight \\
     &        &        & (km\,s$^{-1}$) & (km\,s$^{-1}$) & \\
\hline
2354.560 & 0.542 & B &  $-11.7$ &  $-11.7$ & 0.1 \\
2355.600 & 0.825 & V &   290.7  & $-319.3$ & 0.7 \\
3032.716 & 0.074 & B & $-131.9$ &   182.9  & 0.2 \\
3032.742 & 0.081 & V &     9.1  &     9.1  & 0.1 \\
3033.715 & 0.346 & V & $-288.4$ &   277.4  & 0.7 \\
3033.739 & 0.353 & B & $-285.1$ &   292.6  & 1.0 \\
3034.715 & 0.618 & V &   235.8  & $-219.1$ & 0.7 \\
3034.741 & 0.625 & B &   287.6  & $-265.5$ & 1.0 \\
3035.719 & 0.892 & V &   269.3  & $-222.7$ & 0.7 \\
3035.746 & 0.899 & B &   245.2  & $-219.4$ & 1.0 \\
3036.719 & 0.164 & V & $-259.7$ &   287.0  & 0.7 \\
3036.743 & 0.170 & B & $-302.6$ &   309.5  & 1.0 \\
3037.738 & 0.441 & V &  $-78.3$ &  $-78.3$ & 0.1 \\
3037.762 & 0.447 & B &  $-74.0$ &  $-74.0$ & 0.1 \\
3038.723 & 0.709 & V &   363.5  & $-354.0$ & 0.7 \\
3038.746 & 0.715 & B &   374.6  & $-375.9$ & 1.0 \\
3039.718 & 0.980 & V &    12.4  &    12.4  & 0.1 \\
3039.752 & 0.989 & B &  $-28.0$ &  $-28.0$ & 0.1 \\
3040.717 & 0.252 & V & $-309.3$ &   368.4  & 0.7 \\
3040.741 & 0.258 & B & $-336.1$ &   394.0  & 1.0 \\
3041.722 & 0.525 & V &     4.3  &     4.3  & 0.1 \\
3041.747 & 0.532 & B &    22.9  &    22.9  & 0.1 \\
3042.714 & 0.795 & B &   375.4  & $-374.4$ & 0.2 \\
\hline
\end{tabular}
\end{table}

\begin{figure}[htb]
\begin{center}
\resizebox{8cm}{6.3cm}{\includegraphics{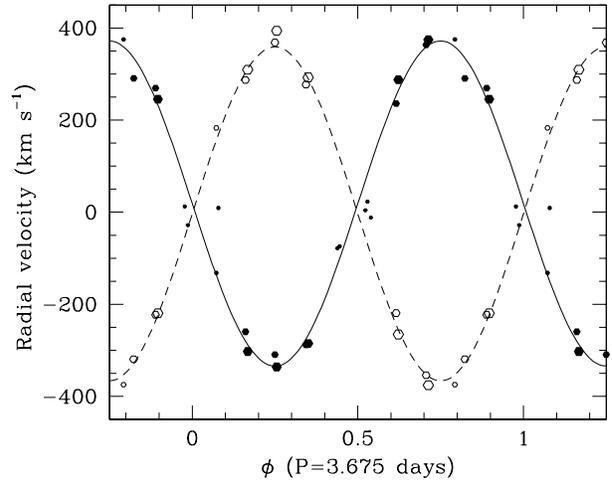}}
\end{center}
\caption{Radial velocity curve of WR\,20a for an orbital period of 3.675\,days. The filled (resp.\ open) symbols stand for the primary (resp.\ secondary) RVs. The sizes of the symbols reflect the relative weight assigned to the various points in the orbital solution (see Table\,\ref{journal}). The solid line and the dashed line yield our best fit orbital solution assuming a circular orbit (see Table\,\ref{tab-orbit}).\label{phase}}
\end{figure}

Due to the ambiguity about the right alias in the periodogram, we caution that the orbital phases as well as the primary/secondary identifications in Table\,\ref{journal} are only valid if the orbital period is indeed $3.675$\,days. Adopting this period, we have derived the orbital solution of WR\,20a using the technique described by Sana et al.\ (\cite{Sana}). Since the RVs show no indication of a significant orbital eccentricity, we assumed a circular orbit. The results are provided in Table\,\ref{tab-orbit} and the RV curve is illustrated in Fig.\,\ref{phase}. We emphasize that orbital solutions derived from the RVs of individual spectral lines overlap within the errors with the results in Table\,\ref{tab-orbit}.
\begin{table}[htb]  
\caption{\label{tab-orbit} Orbital solution for WR\,20a assuming a circular orbit and an orbital period of $3.675$\,days. T$_0$ refers to the time of conjunction with the primary being behind. $\gamma$, $K$ and $a\,\sin{i}$ denote respectively the systemic velocity, the amplitude of the radial velocity curve and the projected separation between the centre of the star and the centre of mass of the binary system. R$_{\rm RL}$ stands for the radius of a sphere with a volume equal to that of the Roche lobe computed according to the formula of Eggleton (\cite{Egg}). All error bars indicate 1-$\sigma$ uncertainties.}
\begin{center}
\begin{tabular}{l r r}
\hline\hline
 & Primary & Secondary\\
\hline
T$_0$(HJD$-2\,450\,000$) & \multicolumn{2}{c}{$3043.480 \pm 0.014$} \\
$\gamma$ (km\,s$^{-1}$) & $18.8 \pm 5.6$ & $-3.5 \pm 5.7$ \\
$K$ (km\,s$^{-1}$) & $353.1 \pm 8.6$ & $362.6 \pm 8.8$ \\
$a\sin i$ (R$_{\odot}$) & $25.6 \pm 0.2$ & $26.3 \pm 0.6$ \\ 
$q = m_1/m_2$ & \multicolumn{2}{c}{$1.03 \pm 0.03$} \\
$m\sin^3 i$ (M$_{\odot}$) & $70.7 \pm 4.0$ & $68.8 \pm 3.8$ \\
R$_{\rm RL}\,\sin{i}$ (R$_{\odot}$) & $19.8 \pm 0.2$ & $19.5 \pm 0.2$ \\ 
\hline
\end{tabular}
\end{center}
\end{table}

The most important results are the large minimum masses of $68.8 \pm 3.8$ and $70.7 \pm 4.0$\,M$_{\odot}$ derived for the secondary and primary component respectively. It should be emphasized that these values strongly depend on the adopted orbital period. In fact, the alternative orbital period ($4.419$\,days) would lead to even larger minimum masses of $80.8 \pm 5.1$ and $84.0 \pm 5.3$\,M$_{\odot}$ respectively. If, on the contrary, the $1 + \nu_1$ alias corresponding to an orbital period of $1.293$\,days were selected, the minimum masses would be much lower ($24.9 \pm 1.4$ and $25.9 \pm 1.4$\,M$_{\odot}$). We note however, that in the latter case, the orbital solution is of significantly lower quality than the one obtained for an orbital period of $3.675$\,days. 
\section{Discussion and future perspectives \label{discuss}}
The WN6ha or O3\,If$^*$/WN6ha spectral type of the components of WR\,20a is significantly earlier than those of the earliest members (O6:V -- O7:V) of Westerlund\,2 analysed by Moffat et al.\ (\cite{Moffat}). 

Several authors have estimated the distance of Westerlund\,2: $7.9^{+1.2}_{-1.0}$\,kpc (Moffat et al.\ \cite{Moffat}); $5.7 \pm 0.3$\,kpc (Piatti et al.\ \cite{Piatti});  $6.4 \pm 0.4$\,kpc (Carraro \& Munari \cite{CM}). Adopting a visual brightness ratio of 1 together with the photometry of Moffat et al.\ (\cite{Moffat}; $V = 13.58$, $B - V = 1.51$) the various distance estimates yield absolute magnitudes $M_{\mathrm V}$ in the range $-5.0$ to $-6.1$. This is fainter than the average of three WN6ha stars in NGC\,3603, namely $-6.9$ to $-7.7$\,mag\footnote{Moffat et al.\ (\cite{MDS}) derived an average $M_{\mathrm V}$ of $-7.7$ for the three H-rich WN stars in NGC\,3603, whilst Crowther \& Dessart (\cite{CD}) re-evaluated their distance and reddening and obtained an average of $-6.9$\,mag.}, such that only the largest published distance of Westerlund\,2 would be in marginal agreement with the photometry of WR\,20a, allowing for a cosmic scatter of 0.5 -- 0.7\,mag on $M_{\mathrm V}$. This problem would be less severe if the absolute magnitudes of the components of WR\,20a were more typical of O3\,I stars ($M_{\mathrm V} = -5.9$ according to Crowther \& Dessart \cite{CD}). Alternatively, WR\,20a could actually be unrelated to Westerlund\,2 (though the rarity of early-type stars and the angular proximity of WR\,20a to the cluster strongly suggest membership) or the photometry of Moffat et al.\ could be affected by photometric eclipses. Concerning the last possibility, we note that given the extremely large minimum masses of the components in WR\,20a, it appears very likely that the system could display eclipses. A photometric monitoring of this binary is therefore of the utmost importance in order to help derive its absolute magnitude, to get rid of the aliasing problem of the RV data and to constrain the orbital inclination. We intend to organize such a campaign in the near future.

An important issue in the context of the formation and the evolution of very massive stars is that of the mass of the most massive ones. Masses up to $\sim 200$\,M$_{\odot}$ have been inferred for the most luminous O2 stars from a comparison with evolutionary tracks (Walborn et al.\ \cite{WHL02}). However, to date, none of the O2 stars is known to belong to a binary system that would allow to determine its mass in a less model dependent way. The most massive main-sequence stars in binary systems known so far were found in R\,136-38 (O3\,V + O6\,V) and R\,136-42 (O3\,V + O3\,V). Massey et al.\ (\cite{MPV}) determined masses of 57 and 40\,M$_{\odot}$ respectively for the O3\,V primary stars of these systems. Another extremely massive object is the WN7ha primary in the 80-day period binary WR\,22 (Rauw et al.\ \cite{RV96}; Schweickhardt et al.\ \cite{SSS99}) with a minimum mass of 55\,M$_{\odot}$. WR\,20a thus probably consists of two of the most massive stars with a direct mass determination known so far. Therefore, WR\,20a is certainly a cornerstone system for future investigations of massive star evolution.

Finally, we note that the emission lines in the spectrum of WR\,20a display strong phase-locked profile variability. This aspect will be discussed along with an analysis of the spectrum with a model atmosphere code in a forthcoming paper.   

\acknowledgement{The Li\`ege team is greatly indebted to the Fonds National de la Recherche Scientifique (Belgium) for multiple assistance. This research is also supported in part by contract P5/36 ``P\^ole d'Attraction Interuniversitaire'' (Belgian Federal Science Policy Office) and through the PRODEX XMM-OM and INTEGRAL projects. The SIMBAD database has been consulted for the bibliography.}

\end{document}